\documentclass[twocolumn,showpacs,amsmath,amssymb,prd,nofootinbib]{revtex4-2}
\usepackage{graphicx}
\usepackage{epsfig}
\usepackage{enumerate}
\def\beq{\begin{equation}}
\def\eeqno#1{\label{#1}\end{equation}}

\def\rarrow{\rightarrow }

\def\dleft{\rlap{{\it D}}\raise 8pt
\hbox{$\scriptscriptstyle\Leftarrow$}}
\def\dright{\rlap{{\it
D}}\raise 8pt\hbox{$\scriptscriptstyle\Rightarrow$}}

\def\az{a_{0}}

\def\l0{\ell_{0}}

\def\rar{\rightarrow}
\def\s{\sigma}

\def\b{\beta}
\def\c{\gamma}
\def\l{\lambda}

\def\f{\phi}
\def\t{\theta}

\def\r{\rho}

\def\n{\nu}
\def\z{\zeta}

\def\av#1{\langle#1\rangle}

\def\A{\mathcal{A}}

\def\SS{\mathcal{S}}
\def\o{\omega}

\def\d{\delta}

\def\xlimin{{x\rarrow\infty \atop{\raise 1pt\hbox to 30pt
{\rightarrowfill}}}}
\def\limlim#1#2{{#1\rarrow #2 \atop{\raise 1pt\hbox to 30pt
{\rightarrowfill}}}}

\def\vr{{\bf r}}

\def\vv{{\bf v}}

\def\S{\Sigma}

\def\div{\vec \nabla\cdot}

\def\gN{g\_N}

\def\b{\beta}

\def\n{\nu}

\def\K{\mathcal{K}}

\def\_#1{_{\scriptscriptstyle #1}}
\def\^#1{^{\scriptscriptstyle #1}}

\def\azg{\A_0}

\def\yout{y\_o}
\begin{document}
\title{Deep-MOND polytropes}

\author{Mordehai Milgrom}
\affiliation{Department of Particle Physics and Astrophysics, Weizmann Institute}

\begin{abstract}
Working within the very-low-acceleration limit of MOND -- the deep-MOND limit (DML), I describe spherical, self-gravitating systems governed by a polytropic equation of state, $P=\K\r^\c$ ($P$ is the pressure, and $\r$ the density). As self-consistent structures, such idealized systems can serve as heuristic models for DML, astronomical systems, such as dwarf spheroidal galaxies, low-surface-density elliptical galaxies and star clusters, and diffuse galaxy groups. They can also serve as testing ground for various theoretical MOND inferences. In dimensionless form, the equation satisfied by the radial density profile $\z(y)$ is (for $\c\not=1$)
$[\int_0^y \z \bar y^2 d\bar y]^{1/2}=-yd(\z^{\c-1})/dy$.
Or, $\t^n(y)=y^{-2}[(y\t')^2]'$,
where $\t=\z^{\c-1}$, and $n\equiv (\c-1)^{-1}$.
I discuss general properties of the solutions, contrasting them with those of their Newtonian analogues -- the Lane-Emden polytropes. Due to the stronger MOND gravity, all DML polytropes have a finite mass, and for $n<\infty$ ($\c>1$) all have a finite radius (Lane-Emden spheres have a finite mass only for $n\le 5$).
I use the DML polytropes to study DML scaling relations. For example, they satisfy a universal relation (for all $\K$ and $\c$) between the total mass, $M$, and the mass-average velocity dispersion $\s$: $MG\az=(9/4)\s^4$. However, the relation between $M$ and other measures of the velocity dispersion, such as the central, projected one, $\bar\s$, does depend on $n$ (but not $\K$), defining a `fundamental surface' in the $[M,~\bar\s,~n]$ space. I also describe the generalization to anisotropic polytropes, which also all have a finite radius (for $\c>1$) and which all satisfy the above universal $M-\s$ relation. This more extended class of models exhibits the central-surface-densities relation: a tight relation between the baryonic and the dynamical central-surface-densities predicted by MOND.
\end{abstract}
\maketitle

\section{Introduction}
MOND \cite{milgrom83} is a theory of dynamics that strives to supplant Newtonian dynamics and general relativity. Its main motivation is to account for the dynamics of galactic systems and the Universe at large without the `dark' constituents that are required in the standard dynamics. Extensive reviews of MOND can be found in references
\cite{fm12}, \cite{milgrom14}, and \cite{milgrom20}.
\par
MOND introduces a new constant, $\az$, with the dimensions of acceleration. It reduces to standard dynamics when all accelerations in the system at hand are much higher than $\az$. In the opposite limit -- the deep-MOND limit (DML) -- when all relevant accelerations are much below $\az$, dynamics become space-time scale invariant \cite{milgrom09}.
\par
Given a theory, it is helpful to construct heuristic models of systems thought to be governed by the theory, either as approximate descriptions of actual systems, or as testbeds for studying various aspects of the dynamical theory at hand.
In the case of MOND, the relevant systems are stellar and galactic systems -- such as star clusters, disc and spheroidal galaxies, clusters of galaxies, etc.
\par
Examples of heuristic models for disc galaxies are the Kuzmin and the Mestel discs (e.g., Ref. \cite{bt08}).
\par
In standard dynamics, heuristic models for spheroidal astronomical objects have also been in use for many years. Notable among them are the Lane-Emden spheres (described, e.g., in Refs. \cite{bt08,kippen12}), which are self-gravitating spheres made of matter with isotropic velocity distribution that satisfies a polytropic equation of state (EoS) -- a relation between pressure, $P$, and density, $\r$, of the form
\beq P=\K\r^{\c}.  \eeqno{polit}

\par
Examples of polytropes in astrophysics are zero-temperature white dwarfs and neutron stars, where $\c$ can vary between that of a relativistic gas ($\c=4/3$) to that of nonrelativistic gas ($\c=5/3$) (e.g.,  Ref. \cite{kippen12}).
\par
As stellar systems, such polytropes arise as spheres with isotropic velocity dispersions, with a phase-space distribution function, $f$, that depends on velocity and position only through the energy, $f(\vr, \vv)=f(E)$, where $E=\vv^2/2+\f$ ($\f$ is the gravitational potential), and, furthermore $f(E)\propto (-E)^{n-3/2}$ for bound constituents (and 0 for unbound values). This gives a polytrope with $\c=1+1/n$ \cite{bt08}. For such stellar systems, Ref. \cite{bt08} shows that one must have $n>1/2$ ($\c<3$), but $n\le 1/2$ is allowed in general, and I shall consider all values of $n$ between 0 and $\infty$ (inclusive).
\par
In the context of MOND, the first class of heuristic, spherical models to be discussed  -- `isothermal spheres' -- were described in detail in Ref. \cite{milgrom84}. These are self-gravitating spheres with a radius-independent, but possibly anisotropic velocity distribution. Their isotropic version is a subclass of polytropic spheres, with $\c=1$.
\par
More general MOND polytropes (with a polytropic EoS, but obeying MOND dynamics) have been considered as models for astrophysical systems.
For example, Sanders, in Ref. \cite{sanders00}, considered MOND polytropic spheres as models of elliptical galaxies. Beyond the polytropic EoS, Sanders allowed anisotropic velocity distributions, with radius dependent degree of anisotropy.
He considered correlations between global properties of these models -- such as mass, size, and velocity dispersions (without discussing the structure of spheres). He showed that the larger variety afforded by anisotropic polytropes -- beyond that of MOND isothermal spheres -- is needed to make these spheres close models of observed elliptical galaxies. For example, he showed that near-isothermal polytropes -- those with $\c$ values near 1 -- better account for the so-called `fundamental plane' of ellipticals -- an observed correlation involving global properties.
\par
Later on, Refs. \cite{ibata11} and \cite{sanders12} debated the adequacy of MOND polytropes for describing the structure and dynamics of one specific globular cluster, NGC 2419.\footnote{ Reference \cite{sanders12} even allowed for a radius-dependent polytropic index.}
\par
Both the `high-surface-density' ellipticals considered in Ref. \cite{sanders00}, and typical globular clusters (including NGC 2419), are characterized by high accelerations ($g\gg\az$) in their main body, and dip into the DML only at their outskirts. Their description in MOND thus requires accounting for the full gamut of accelerations between the Newtonian and the DML extremes.
\par
But here I describe and discuss mainly the more limited subclass of  deep-MOND polytropes (DMPs), isotropic and anisotropic. While this class of models is more limited in scope, it affords concentrating in more detail on the properties of such spheres.
And to boot, it does already capture many of the idiosyncracies of MOND that are also shared by the wider classes. Some of these characteristics, on the other hand, are specific to the DML and its scale invariance.
\par
DMPs can serve as heuristic models for actual DML astrophysical systems -- such as dwarf spheroidal galaxies and galaxy groups -- if not in their details, at least in the general scaling relations that these models obey, and which I discuss.
For example -- as I discuss in some detail here -- they show the strong correlation between the central-surface-densities of baryons and of the putative dark matter halo, pointed to and considered for galaxies in Refs. \cite{milgrom09a,lelli16,milgrom16}.
\par
They are also used here, with several examples, to demonstrate that the DML $M-\s$ relation is not unique, but may depend on the particular choice of $\s$ measure.
\par
In Sec. \ref{equation}, I derive the deep-MOND polytropic equation in its different forms. In Sec. \ref{solutions}, I describe properties of the spheres: the runs of density, accumulated mass, etc. and the structure near their center and near their boundary.  Section  \ref{scaling} discusses scaling relations among local and global properties. Enlarging the scope, I describe in Sec. \ref{anisotropic} DMPs with a constant velocity anisotropy. Section \ref{discussion} is a discussion.

\section{Deep-MOND polytropic equation \label{equation}}
`Modified-gravity' formulations of MOND -- such as the `aquadratic Lagrangian' formulation \cite{bm84}, and the `quasi-linear' formulation \cite{milgrom10} -- predict that in spherical systems, such as I discuss here,  the acceleration, $g$, at radius $r$, is given by
\beq  g=\n(\gN/\az)\gN,  \eeqno{meq}
where $\gN=M(r)G/r^2$ is the Newtonian acceleration,
and $\n$ is the MOND interpolating function having the following behavior: In the high-acceleration, Newtonian regime, $\gN\gg\az$, $\n(q\gg 1)\approx 1$, and in the DML, we have $\n(q\ll 1)\approx q^{-1/2}$. This functional dependence of $g$ on $\gN$ coincides with the original formulation of MOND in Ref. \cite{milgrom83}.
\par
The hydrostatic-equilibrium equation, which determines the structure of self-gravitating spheres with isotropic velocity distribution, is
\beq \r g=-\frac{dP}{dr}. \eeqno{equil}
\par
In the DML we can thus write
\beq \r\left[\frac{\azg M(r)}{r^2}\right]^{1/2}=-\frac{dP}{dr},  \eeqno{equilI}
where $M(r)=4\pi\int_0^r \r(\bar r)\bar r^2~ d\bar r$, is the accumulated mass, and $\azg\equiv \az G$ is the `second MOND constant', and is the only combination of $\az$ and $G$ that can appear in DML equations, as a result of the scale invariance of this limit \cite{milgrom09}.
\par
Consider self-gravitating spheres whose constituents satisfy a polytropic EoS, and which have an isotropic velocity dispersion,
\beq \s^2\equiv 3P/\r=3\K\r^{1/n}, \eeqno{dda}
where
\beq n\equiv \frac{1}{\c-1},   \eeqno{gaman}
is used alternately with $\c$.
\par
Substituting the EoS in eq. (\ref{equilI}), we get the integro-differential equation for the density run of DML polytropes (for regions where $\r\not =0$)\footnote{$\r=0$ is always a solution of eq. (\ref{equil}) for a polytropic EoS.} . For $\c\not=1$,
\beq \left[\int_0^r \r(\bar r)\bar r^2~ d\bar r\right]^{1/2}=-(n+1) Sr\frac{d\r^{1/n}}{dr},\eeqno{juta}
where
 \beq S\equiv \frac{\K}{\sqrt{4\pi\azg}}. \eeqno{sasa}
For the special case $\c=1$ ($n=\infty$), the equation of hydrostatic-equilibrium is
\beq \left[\int_0^r \r(\bar r)\bar r^2~d\bar r\right]^{1/2}=-S\frac{d\ln\r}{d\ln r}.\eeqno{jutama}
[This can also be obtained by taking the limit $n\rar\infty$ of eq. (\ref{juta}), since to order $1/n$, $\r^{1/n}=1+n^{-1}\ln\r+O(n^{-2})$.]
\par
Equation (\ref{juta}) is to be contrasted with the Newtonian polytrope equation (for $\c\not=1$)
 \beq \int_0^r \r(\bar r)\bar r^2~ d\bar r=-(n+1)S\_N r^2\frac{d\r^{1/n}}{dr},\eeqno{jutaha}
where
 \beq S\_N\equiv \frac{1}{4\pi}\frac{\K}{G}. \eeqno{samar}

\subsection{Dimensional analysis and dimensionless forms}
The coefficient $S$ in eq. (\ref{sasa}) has dimensions $[S]=[\ell^{3/n}][M^{1/2-1/n}]$.
We would like to absorb it by defining length and mass units and express all quantities in these units.
\par
Two special values of $\c$ suggest themselves. For $\c=1$  ($n=\infty$), $S$ has dimensions of $[M^{1/2}]$, and $\K$ has dimensions of velocity squared. This is the isothermal-sphere case -- discussed in detail in Ref. \cite{milgrom84} -- where the constant velocity dispersion
($\s^4 \propto \K^2$) determines the total mass $M=(81/4)(\K^2/\azg$) independent of the size. Scaling up the size of any such system, gives another sphere with the same velocities and total mass, but scaled density.
\par
The other special case is $n=2$, where the dimensions of $S$ are $[\ell^{3/2}]$. The polytropic coefficient
then defines an absolute size scale for the system  $\ell\equiv S^{2/3}$. Systems with $n=2$ and a given $\K$ have all the same size, but different masses (and thus densities).
In terms of the dimensionless radius $w\equiv r/9^{1/3}\ell$ we write eq. (\ref{juta}) for this case as
\beq \left[\int_0^w \r(\bar w)\bar w^2~ d\bar w\right]^{1/2}=-w\frac{d\r^{1/2}}{dw}.\eeqno{jupata}
Clearly, if $\r(w)$ is a solution, so is $a\r(w)$ for any $a$.
\par
For other values of $n$, we can choose a length unit, and $S$ then defines mass and density scales,
with which we construct the dimensionless form of the equations. The values of $n$ and $\K$
 define a  Mass-radius relation.
\par
As somewhat of an aside, note that,
in principle, MOND involves two dimensionful constants, $G$ and $\az$, which together with $\K$ can be used to define absolute, `natural' scales of length, mass, and density. For example, the relation $\hat MG/\ell=\az$ could be used together with $S=\ell^{3/n}\hat M^{1/2-1/n}$ to define the scales $\hat M$ and $\ell$ (and from them construct a density scale).
\par
This procedure is, however, neither useful nor desirable for systems in the present DML context, which is the limit $\az\rar\infty$, $G\rar 0$, where it is `illegal' to use $G$ and/or $\az$ separately. Only the combination $\azg=G\az$ can be used. Viewed differently, the inutility of such a procedure can be seen by noting that DML systems have typical sizes that are much larger
(infinitely in the limit) than their MOND radius defined as $r\_M\equiv MG/\az$. So it is not instructive to measure lengths in units of the MOND radius.
\par
So, we proceed by defining some arbitrary length unit $\ell$ and define a density unit $\hat \r$
such that
\beq \ell^3\hat \r^{1-2/n}=(n+1)^2S^2.  \eeqno{jayu}
Then, defining $\z\equiv \r/\hat\r$ and $y\equiv r/\ell$, we write eq. (\ref{juta}) as the deep-MOND, dimensionless, polytropic integro-differential equation for $\z(y)$
\beq m^{1/2}(y)\equiv \left[\int_0^y \z \bar y^2 d\bar y\right]^{1/2}=-y\frac{d\z^{1/n}}{dy},  \eeqno{dimless}
where $m(y)$ is the dimensionless accumulated mass.
As in the Newtonian limit, it is useful to work with the dependent variable $\t=\z^{1/n}$, in terms of which
\beq m^{1/2}(y)\equiv\left[ \int_0^y \t^n \bar y^2 d\bar y\right]^{1/2}=-y\t'.  \eeqno{dimrat}
Or, converted into a 2nd order differential equation, we have
\beq \t^n(y)=y^{-2}[(y\t')^2]'=2\t'\t''+\frac{2}{y}(\t')^2.  \eeqno{dimar}
\par
The Newtonian analogue, dimensionless Lane-Emden equation reads
\beq \t^n(z)=-z^{-2}(z^2\t')'=-\t''-\frac{2}{z}\t'.  \eeqno{lane}
The dimensionless variables are defined differently, and the equation is of a very different nature (e.g., the right-hand side of the latter is linear in $\t$).
\par
We see from eq. (\ref{dimrat}) that hydrostatic-equilibrium with the EoS dictates that inside the sphere, where $\z>0$, the quantity $\t\equiv \z^{1/n}$ satisfies $\t'=-m^{1/2}/y$, which is the dimensionless, DML acceleration ($-m/y^2$ being the dimensionless Newtonian acceleration).  Thus, inside the sphere (but not outside) $\t=-\f$, where $\f$ is the dimensionless DML gravitational potential, with its free additive constant chosen such that $\f$ vanishes where the density does, which in our case is at the edge of the sphere (see below).
\par
Equation (\ref{dimar}) requires specifying two boundary conditions.
We shall see below that we need to have $\t'(0)=0$. The value $\t(0)$ can take up arbitrary positive values, but the solutions for different $\t(0)$ values are simply related to each other. Equation (\ref{dimar}) is invariant to $\t\rar \l\t$ and $y\rar \l^{(2-n)/3}y$. So the solution for one value of $\t(0)$ generates those for all other values.
I thus set $\t(0)=1$ hereafter.
\par
Once the solution $\t(y)$ is found for a given $n$ and $\t(0)=1$, call it $\t\_{(n)}(y)$, the solution for any value of $\t(0)$ is
\beq \t(y)=\l\t\_{(n)}[\l^{-(2-n)/3}y];~~~~~\l=\t(0).   \eeqno{tutaz}
\par
Putting together all the reductions we made, one finds that for given $n$ and $\K$, the solution $\t\_{(n)}(y)$ fans into a one-parameter family of solutions. The parameter spanning this family can be the unit length $\ell$ chosen arbitrarily, or the unit density $\hat\r$.
\par
One useful choice of $\hat\r$ is the central density $\r_0\equiv\r(0)$, in terms of which the general solution is
\beq \r(r)=\r_0 \t\_{(n)}^n[\r_0^{(n-2)/3n}(n+1)^{-2/3}S^{-2/3}r].   \eeqno{gabul}
\par
We shall also use below an alternative independent variable defined by
\beq x= \frac{2}{\sqrt{27}}y^{3/2}.   \eeqno{kapa}
Then, eq. (\ref{dimar}) becomes
\beq \t^n(x)=\frac{1}{2x}[(x\t')^2]'=x\t'\t''+(\t')^2,  \eeqno{dimush}
where here the derivatives are with respect to $x$.
\par
In the Newtonian case, the dimensions of $S\_N$ are $[\ell^{3/n-1}][M^{1-1/n}]$. One special value of $n$ is
$n=3$ ($\c=4/3$), where, in analogy with the DML $n=\infty$, isothermal case, the mass is determined by the value of $\K$. This value of $n$ occurs, e.g., in the EoS of a relativistic degenerate fermion gas, where the special value of the mass determines the maximal mass of a white dwarf.
\par
The other special value is $n=1$ ($\c=2$), where the dimensions of $S\_N$ are $[\ell^2]$, resulting in a size of the polytrope that -- in analogy with the DML $n=2$ case -- is determined by $\K$.
\subsection{Loss of scale invariance  \label{losssi}}
One of the basic tenets of MOND is that the DML is scale invariant \cite{milgrom09}, in the sense that the equations describing self gravitating systems in the DML are invariant to space-time scaling of the system's degrees of freedom, $(r,t)\rar \l(r,t)$. Since we are dealing here with time independent systems, scale invariance implies invariance to scaling of all lengths, $r \rar \l r$. Such invariance means that scaling all lengths in a solution of the theory produces another solution.
\par
It may thus appear puzzling that for the general case, $\c\not= 1$, our eq. (\ref{juta}), which is meant to describe self-gravitating systems, is not scale invariant: Under scaling (under which $\r\rar\l^{-3}\r$) the left-hand side is invariant, but the right-hand side is scaled by $\l^{-3/n}$.
\par
The scale invariance of DML gravity is encapsulated in the fact that the only dimensionful constant that appears in it is $\azg$, whose dimensions, $[\azg]=[\ell^4][t^{-4}][M^{-1}]$, are invariant to scaling {\it of length and time units}. However, here, by imposing the polytropic EoS, we introduce another dimensionful constant,
$\K$, whose dimensions are $[\K]=[\ell^{2+3/n}][t^{-2}][M^{-1/n}]$, and whose value (for $n<\infty$) does change under scaling of the time-length units.
So, if a given sphere is a solution for a certain $\K$, the scaled sphere will be a solution, not for the same $\K$, but  for $\K'= \l^{3/n}\K$.
\par
When the EoS results from microscopic considerations, as in the EoS of a degenerate fermion gas, the problem is no more one of pure gravity. The microscopic physics breaks down the scale invariance: It introduces dimensionful constants that are not invariant to unit scaling (in contract with $\azg$). The constant $\K$ is then a `constant of nature'.
\par
However, what is the underlying cause for scale-invariance breakdown in the application to galactic systems, which are purely gravitational? The cause is hidden in the processes that supposedly drive such self gravitating systems to become approximate polytropes. We do not know what such processes are, and so we do not know why they should cause the polytropic coefficient to depend on system size, as described above, thus apparently breaking scale invariance. But this need not worry us further.

\section{Deep-MOND polytropes: Solutions \label{solutions}}
\subsection{Analytic solutions \label{analytic}}
For the Newtonian, Lane-Emden equation, analytic solutions are known for $n=0,~1,~5$ \cite{kippen12}.
The case $n=1$ gives a linear equation and is easy to solve analytically [the solution is $\t=\sin(y)/y$].
\par
For  DMPs, I have so far identified analytic solutions for two $n$ values: The isothermal case, $\c=1$ discussed in detail in Ref. \cite{milgrom84}, where it was found that\footnote{The actual analytic solution was found at the time by Israel Kovner, private communication.}
\beq \r\_{(\infty)}(y)=a[1+(y/\ell)^{3/2}]^{-3},   \eeqno{uta}
where $\ell$ is arbitrary, and $a$ is determined from $\ell$ so that the total mass is  $M=(81/4)(\K^2/\azg$), as mentioned above.
\par
This $n=\infty$ case is analogous in some ways to the limiting, $n=5$ case for Lane-Emden which is the largest $n$ for which the mass is finite, which has an infinite extent, and such that all smaller $n$ values correspond to polytropes of a finite radius.
\par
For $n=0$ the solution is
\beq \t\_{(0)}(y)=1-\frac{2}{\sqrt{27}}y^{3/2},   \eeqno{kamul}
or
\beq \t\_{(0)}(x)=1-x,   \eeqno{kalio}
in terms of the variable $x$ defined above eq. (\ref{kapa}).
This corresponds to a constant-density sphere, for which the Newtonian field is harmonic with $\gN\propto r$, hence, the DML field is $g\propto r^{1/2}$, and the DML potential is $\t\propto -r^{3/2}$ up to an additive constant.
\subsection{Numerical solutions \label{numerical}}
\begin{figure}[ht]
	\centering
\includegraphics[width = 7cm] {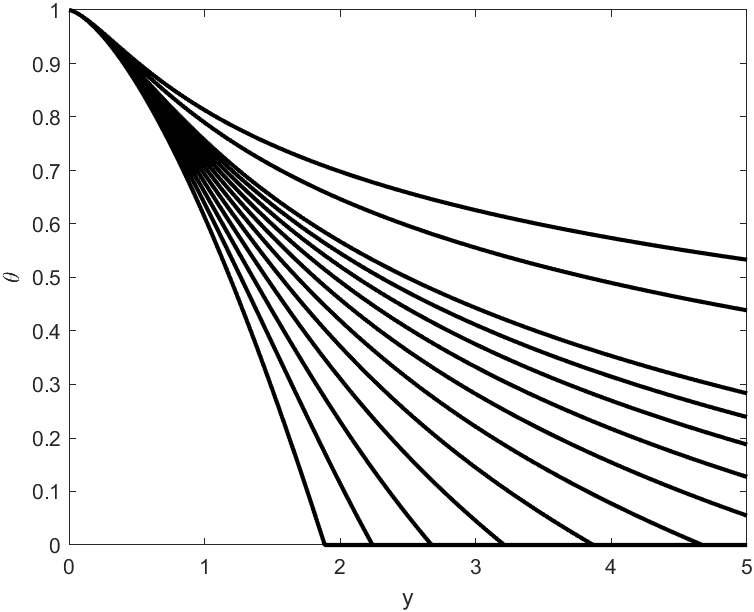}
\caption{The run of $\t\_{(n)}$ [normalized to $\t(0)=1$], for $n=0,~1,~2,~3,~4,~5,~6,~7,~8,~9,~10,~15,~20$ (from bottom to top).}		\label{tplot}
\end{figure}

\begin{figure}[ht]
	\centering
\includegraphics[width = 7cm] {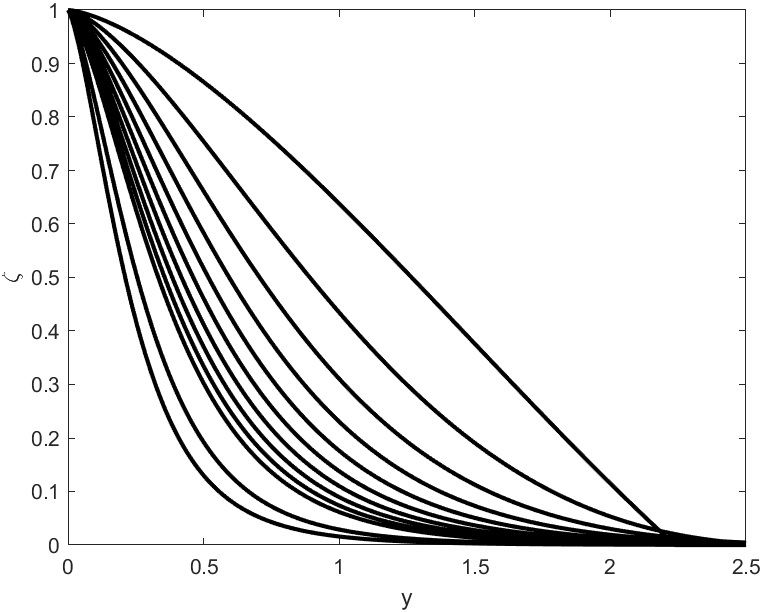}
\caption{ The run of the dimensionless density, $\z(y)$, for $n=1,~2,~3,~4,~5,~6,~7,~8,~9,~10,~15,~20$   (from top to bottom; $n=0$ not shown).     }		\label{rplot}
\end{figure}

\begin{figure}[ht]
	\centering
\includegraphics[width = 7cm] {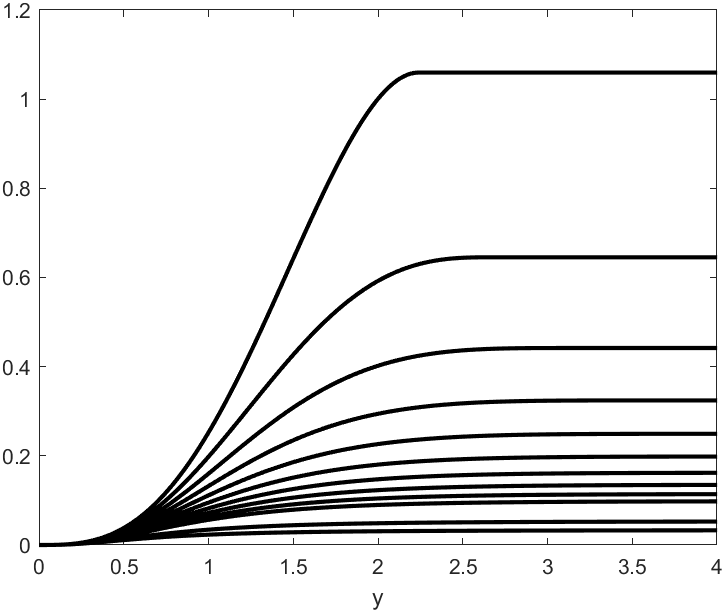}
\caption{  The run of the accumulated dimensionless mass, $m(y)$, for $n=1,~2,~3,~4,~5,~6,~7,~8,~9,~10,~15,~20$ (from top to bottom; $n=0$ not shown). }		\label{mplot}
\end{figure}

\begin{figure}[ht]
	\centering
\includegraphics[width = 7cm] {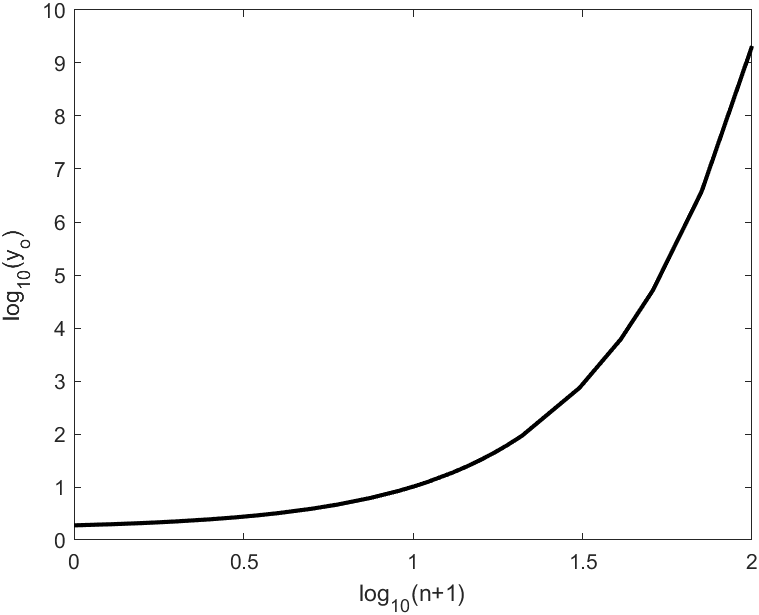}
\caption{The dimensionless edge radius as a function of $n$ for $n\le 99$.}		\label{edgerad}
\end{figure}

In default of more analytic solutions, I now describe the results of solving eq. (\ref{dimar}) numerically.
Figs. \ref{tplot}-\ref{mplot} show the runs of $\t\_{(n)}$, $\z\_{(n)}$, and the accumulated mass for $n=0,~1,~2,~3,~4,~5,~6,~7,~8,~9,~10,~15,~20$.
The corresponding values of the dimensionless edge radii are: $\yout\approx 1.89,~ 2.24,~  2.68,~  3.21,~ 3.87,~  4.67,~  5.66,~ 6.86,~  8.34,~   10.15,$ $12.37,~ 33.73,~ 93.73$, shown in
Fig. \ref{edgerad} as a function of $n$, for $n\le 99$.
The values of the dimensionless masses are, respectively, $m_n\approx 2.24,~ 1.06,~ 0.65,~  0.44,~ 0.32,~  0.25,~ 0.20,~ 0.16,~ 0.13,~ 0.11,$   $0.098,~ 0.053,~  0.033$.
\par
We see that the edge radius increases quickly with $n$, and is $\sim 10^9$ for $n=100$.


\subsection{Behavior around the origin \label{origin}}
If $\r$ diverges at the origin, the right hand of eq.(\ref{juta}), or of eq.(\ref{dimless}), also diverges
(for $\c\not = 1$; $\c=1$ needs a special treatment), which means that the mass diverges there, which we reject. Thus we want solutions with a finite $\r(0)$. With our choice of normalization, this implies $\t(0)=1$. To balance gravity, $P$, hence $\r$, must everywhere decrease outward.
\par
Substituting in eq. (\ref{dimless}) an expansion in $y$ near the origin, one finds, to order $y^3$
\beq \t(y)=1-\frac{2}{\sqrt{27}}y^{3/2}+\frac{2n}{81}y^3+ o(y^3).   \eeqno{kasaba}
For comparison, the behavior of Newtonian polytropes near the origin is \cite{kippen12}
\beq \t\_N(y)=1-\frac{1}{6}y^2+\frac{n}{120}y^4+o(y^4).  \eeqno{uio}
\par
In terms of the variable $x$ defined in eq. (\ref{kapa}), the expansion near the origin to order $x^3$ (order $y^{9/2}$) is

\beq  \t=1-x+\frac{n}{6}x^2+\frac{n}{24}\left(1-\frac{8n}{9}\right) x^3+o(x^4).  \eeqno{mirt}
Here we see the usefulness of using $x$ as independent variable, in terms of which the expansion is simpler, and in particular, $d^2\t/dx^2$ is finite at the origin, while $d^2\t/dy^2$ is infinite.
\par
\begin{figure}[ht]
	\centering
\includegraphics[width = 7cm] {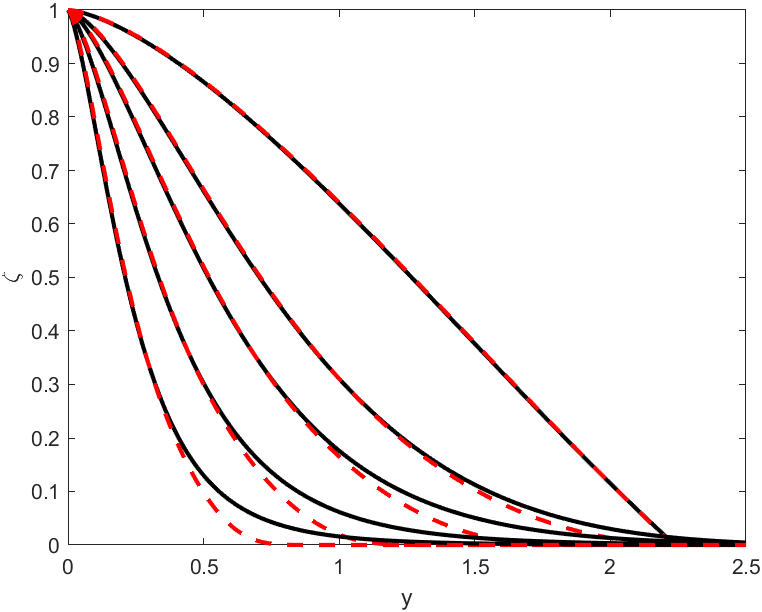}
\caption{Comparison of the dimensionless density, $\z(y)=\t^n(y)$, as given by the expansion series (\ref{mirt}) (red dashed line) with numerical results (black solid line), for $n=1,~3,~5,~10,~20$ (lower $\z$ for higher $n$).}		\label{rhoseries}
\end{figure}

Figure \ref{rhoseries} compares approximation (\ref{mirt}) (expressed as functions of $y$, not $x$) for the density profiles, $\z=\t^n$, with numerical solutions. We see that eq. (\ref{mirt}) is a reasonable approximation in regions where most of the mass is.
(For $n=1$ this approximation is good to better than one percent everywhere in the sphere.)
\par
\begin{figure}[ht]
	\centering
\includegraphics[width = 7cm] {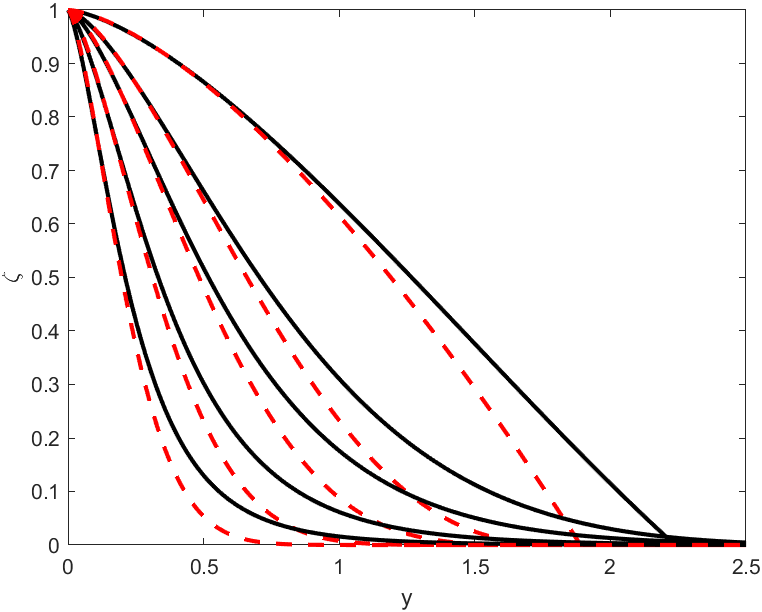}
\caption{Comparison of $\z(y)$, as given by the first two terms in the expansion (\ref{kasaba}) or (\ref{mirt}), $\z\approx (1-x)^n$, (red dashed line) with numerical results (black solid line), for $n=1,~3,~5,~10,~20$ (lower $\z$ for higher $n$).}		\label{rhofirst}
\end{figure}

I also show in Fig. \ref{rhofirst} a comparison of  $\z\approx (1-x)^n$ with the numerical results, because this approximation, while poorer, is useful for various analytic estimates.


\subsection{Behavior at the edge \label{edge}}
I now show that for $\c>1$ ($n<\infty$), DMPs must have a finite edge where the density drops to $0$, and beyond which we discard the solution.
The right-hand side of  eq. (\ref{dimrat}) is $-d\t/d\ln{y}$.
As long as $\t$ is still positive (i.e. within the sphere) it equals $-\f$; so to balance gravity, $\t$ has to be a decreasing function of $y$ (thus of $\ln{y}$). To have the region with $\t>0$ extend to infinity, would mean that $\t(\ln{y})$ is a positive, ever-decreasing function. Such a function must have its derivative tend to 0 at infinity. But, $-d\t/d\ln{y}\rar 0$ at infinity is inconsistent with the left-hand side of eq. (\ref{dimrat}) having to be nondecreasing (and positive). Thus the region where $\t>0$ cannot extend to infinity, and $\t$ must vanish at a finite radius.
\par
This contrasts with the behavior of Newtonian polytropes, which for $n<5$ have a finite radius, for $n=5$ have an infinite extent, but a finite total mass, and have a diverging mass for $n>5$ \cite{kippen12}. The fact that DMPs all have finite masses, and a finite radius for all finite $n$, is due to the stronger MOND gravity.
\par
If $\yout$ is the radius of the edge [where $\t(\yout)=0$], and $m_t\equiv m(\yout)$ is the total mass,
then the dominant behavior just interior to the edge is
\beq \t(y)\approx -m_t^{1/2}\ln\left(\frac{y}{\yout}\right).  \eeqno{logo}
This is because $\ln(y)$ annihilates the right-hand side of eq. (\ref{dimar}), and $\t^n$ on the left-hand side is balanced by higher order terms in $\t$.
\par
\begin{figure}[ht]
	\centering
\includegraphics[width = 7cm] {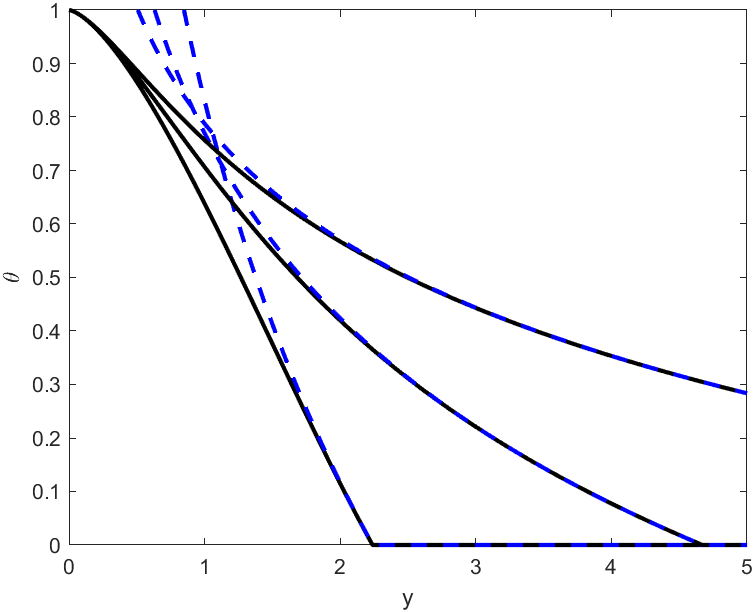}
\caption{The logarithmic approximation to $\t$ from eq.(\ref{logo}), from bottom to top, for $n=1,~5,~10$ (in blue dashed lines), compared with numerical solution (in solid black). }		\label{logonly}
\end{figure}

This logarithmic behavior can be understood as follows: The enclosed mass converges to $m_t$ at radii below $\yout$; for larger $n$ values it does so much below $\yout$ (see Fig. \ref{mplot}). Once this happens, the DML potential becomes that outside a spherical mass $m_t$: $\f=m_t^{1/2}\ln(y)$,
 up to an additive constant, which is fixed by our definition $\t(\yout)=0$ (remember that inside $\yout$, $\t=-\f$).
Figure \ref{logonly} shows directly, for some $n$ values, how far below $\yout$ the logarithmic behavior of eq. (\ref{logo}) is a good approximation.
\par
Our definition $\t\equiv\z^{1/n}$ implies that $\t=0$ outside the edge radius. That $\f=-\t$ inside the sphere follows from the EoS and the dynamics. But, of course, $\f$ does not vanish outside the sphere where it is still given by
\beq \f=m_t^{1/2}\ln(y/\yout). \eeqno{zalu}
\par
The next order correction to $\t$ below the edge gives
\beq \t(y)\approx -m_t^{1/2}\ln\left(\frac{y}{\yout}\right)-b\left(1-\frac{y}{\yout}\right)^{n+2},  \eeqno{logopl}
where
\beq  b=\frac{m_t^{(n-1)/2}\yout^3}{2(n+1)(n+2)}.   \eeqno{juti}
\par
\begin{figure}[ht]
	\centering
\includegraphics[width = 7cm] {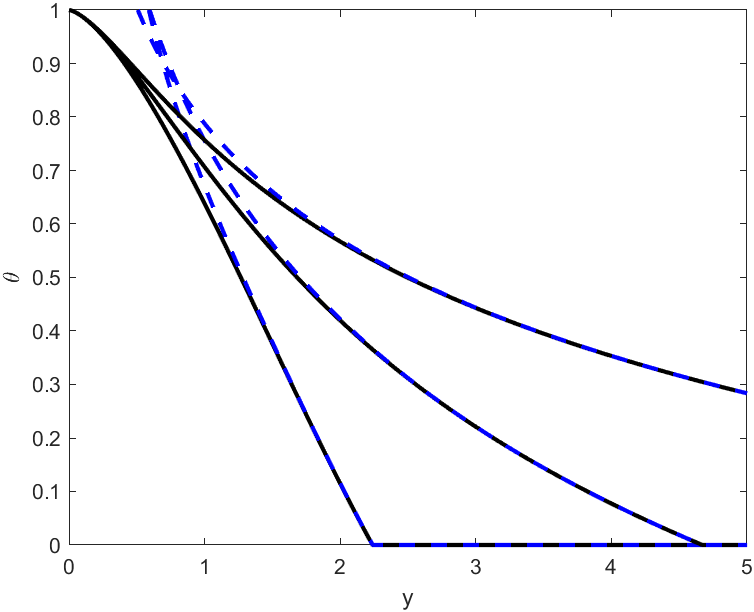}
\caption{The next-to-logarithmic approximation to $\t$ from eqs.(\ref{logopl})(\ref{juti}), from bottom to top, for $n=1,~5,~10$ (in blue dashed lines), compared with numerical solution (in solid black).}		\label{full}
\end{figure}

Numerical solutions for several $n$ values are compared with expression (\ref{logopl}) in Fig. \ref{full}.

\section{Scaling relations  \label{scaling}}
\subsection{Mass-velocity-dispersion relations}
Integrating the density in eq. (\ref{gabul}) to get the total mass, $M$, we find that
\beq M=\frac{(n+1)^2}{9\azg}\s^4(0)\int_0^{\yout}\t\_{(n)}^n(y)y^2~dy=\frac{(n+1)^2}{9\azg}\s^4(0)m_n,  \eeqno{pashu}
where $m_n$ is the total dimensionless mass for $\t\_{(n)}$ (integrated to the edge $\yout$), and $\r_0$ is expressed in terms of the central velocity dispersion $\s(0)$ using expression (\ref{dda}).
\par
\begin{figure}[ht]
	\centering
\includegraphics[width = 7cm] {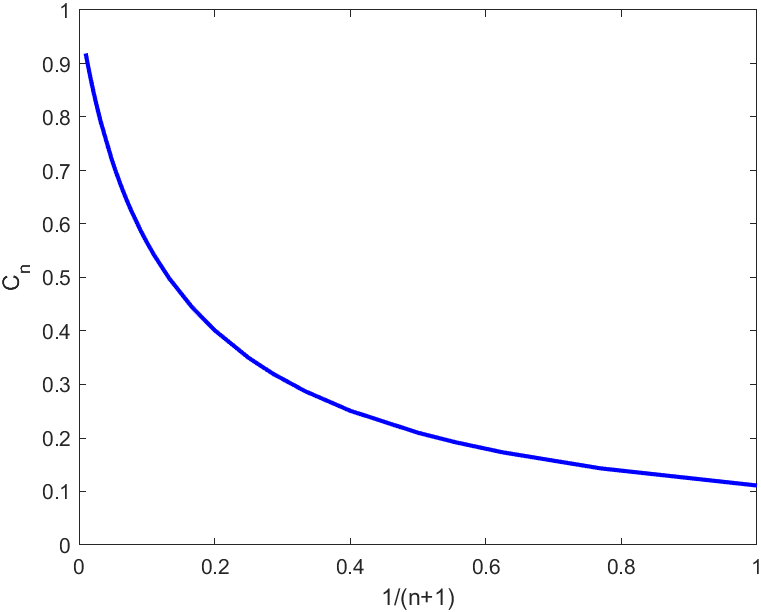}
\caption{The coefficient $C_n\equiv 4(n+1)^2m_n/81$ vs $(n+1)^{-1}$ (for $0\le n\le 100$).}		\label{CN}
\end{figure}

We know from Ref. \cite{milgrom84} that in the isothermal limit, $n\rar\infty$, we have $M\azg=(9/4)\s^4$ [and $\s(0)=\s$].
This means that we have to have in this limit $(n+1)^2m_n\rar 81/4$.
I thus define the coefficients $C_n$ such that eq. (\ref{pashu}) is written as
\beq M\azg=\frac{9}{4}C_n\s^4(0),  \eeqno{pashumm}
where $C_n\equiv 4(n+1)^2m_n/81$ is plotted in Fig. \ref{CN} vs $n$ (one can also read $m_n$ from this figure).
We see that indeed $C_n\rar 1$ for $n\rar \infty$.
$C_n$ varies between 0.1 and 1 for the full range of $n$. But for the higher-$n$ models, which might be more relevant for astrophysical systems, $C_n$ varies only by a factor of about two for all $n\ge 5$.

Thus, DMPs do not share a universal $M-\s(0)$ relation; rather, they span a `fundamental surface' in the three-dimensional parameter space of $[M,~n,~\s(0)]$, described by eq. (\ref{pashumm}). Note that the polytropic coefficient $\K$ does not enter this relation.
\par
More generally, it is easy to see from the dimensional arguments above, that the ratio
$M\azg/\bar\s^4$, for any velocity measure, $\bar\s$, of the spheres, is independent of $\K$, but possibly does depend on $n$.
\par
\begin{figure}[ht]
	\centering
\includegraphics[width = 7cm] {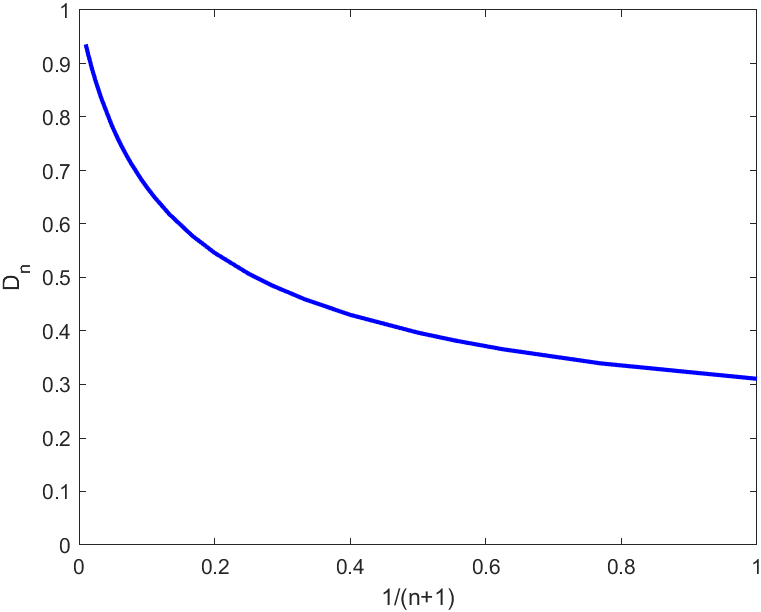}
\caption{The parameter $D_n$, defined in eq. (\ref{tesra}), vs $(n+1)^{-1}$, for $0\le n\le 100$.}
\label{Dn}
\end{figure}

Take, as another example, the more-directly-observed, mass-weighted, line-of-sight, central velocity dispersion, given by
\beq   \s_0^2=\frac{\int\r\s^2dr}{3\int \r~dr}=\frac{\K\int\r^\c dr}{\int\r~dr}. \eeqno{samus}
We can see from the equations relating the density and length scales -- eq. (\ref{jayu}) or eq. (\ref{gabul}) -- that the ratio $M/\s_0^4$ can be gotten from only the dimensionless solution normalized at the origin, $\t\_{(n)}$:
\beq \frac{M\azg}{\s_0^4}=\frac{(n+1)^2(\int y^2\t^n\_{(n)}dy)(\int\t^n\_{(n)}dy)^2}{(\int\t^{n+1}\_{(n)}dy)^2}.  \eeqno{gusgt}
In analogy to $C_n$, define $D_n$ such that
\beq M\azg=\frac{81}{4}D_n\s_0^4,  \eeqno{tesra}
which is shown in Fig. \ref{Dn} as a function of $n$. (The factor of $81$ enters instead of the factor $9$ in eq. (\ref{pashumm}), because here $\s_0$ is the one-dimensional dispersion, whereas $\s(0)$ is a three-dimensional dispersion.) As expected from the results of Ref. \cite{milgrom84}, $D_n\rar 1$ for $n\rar\infty$.

\subsubsection{Universal $M-\av{\s^2}$ relation \label{msigma}}
We saw above that the with some choices of the characteristic velocity dispersion of the polytrope, there exists a triple relation between total mass, $M$, $n$, and $\s$; the ratio $M/\s^4$ is not universal. However, all DMPs -- more generally, all solutions of the DML hydrostatic-equilibrium equation, with any EoS -- do satisfy a universal relation between $M$ and the global, mass-averaged, velocity dispersion.
\beq M\azg=\frac{9}{4}\av{\s^2}^2, \eeqno{poi}
where
\beq \av{\s^2}\equiv M^{-1}\int 4\pi \r(r)\s^2(r)r^2~dr=\frac{12\pi}{M}\int P(r)r^2~dr.  \eeqno{luop}
This holds despite the apparent loss of scale invariance discussed above (but scale invariance of the DML does underlie this relation).
Ref. \cite{gs92}  showed this for the first time for spherical systems. Reference \cite{milgrom94} showed that it holds for arbitrary self-gravitating DML systems in the `aquadratic Lagrangian' theory of Ref. \cite{bm84}, and Ref. \cite{milgrom14a} showed it to be a general result of modified-gravity DML.
\par
In the special case of self-gravitating `gas spheres', this follows from the general DML hydrostatic-equilibrium eq. (\ref{equilI}),
noting that
$$\int P(r)r^2~dr=-\frac{1}{3}\int r^3 P'(r)=\frac{1}{3}\azg^{1/2}\int r^2\r(r)M^{1/2}(r)$$
$$ =\frac{1}{12\pi}\azg^{1/2}\int M'(r)M^{1/2}(r)=\frac{1}{18\pi}\azg^{1/2}\int [M^{3/2}]'(r)$$
\beq = \frac{1}{18\pi}\azg^{1/2}M^{3/2}.  \eeqno{muta}

Putting this together with eq. (\ref{luop}) gives eq. (\ref{poi}).
\subsection{Relations involving the size}
We can also use some measure of the system size as a parameter. This radius will be some function of $n$ because it depends on $\t\_{(n)}$ multiplied by
our length unit
\beq \ell=\r_0^{(2-n)/3n}(n+1)^{2/3}S^{2/3}.  \eeqno{kiol}
\par
For example, we see in Fig. \ref{rhofirst} that approximating $\z$ by the first two terms in the expansion (\ref{kasaba}) $\z(y)\approx [1-(2/\sqrt{27})y^{3/2}]^n$ gives a good approximation up to radii of order $y_h$ -- the radius where $\rho$ drops to half of its central value. So, we can use this approximation to estimate $y_h$ as
\beq y_h\approx \frac{3}{4^{1/3}}(1-2^{-1/n})^{2/3}. \eeqno{halfa}
For $n\gg 1$ this gives to lowest order in $n^{-1}$
\beq y_h\approx \frac{3\ln^{2/3}(2)}{4^{1/3}}n^{-2/3}. \eeqno{halfash}
Since this indicates that $y_h^{3/2}$ decreases as $n^{-1}$, we deduce that approximation (\ref{halfash}) does not improve with increasing $n$, since at $y_h$, higher order terms in the expansion (\ref{kasaba}) do not decrease with $n$.
\par
Approximation (\ref{halfash}) tells us, in conjunction with eq. (\ref{kiol}), that in general, $r_h=\ell y_h$ depends on $n$, $\K$ and $\r_0$.
\par
We can check approximation (\ref{halfash}) by examining the limit of $r_h$  for $n\rar \infty$, and comparing it with the value we can deduce from the results of Ref. \cite{milgrom84} for DML isothermal spheres.
In this limit, the above expression for $r_h$ gives
\beq r_h\rar \frac{3\ln^{2/3}(2)}{4^{1/3}}\r_0^{-1/3}S^{2/3}. \eeqno{halgi}
But, in this limit, $S\rar(M/81\pi)^{1/2}$; so
\beq r_h\rar \frac{3\ln^{2/3}(2)}{(324\pi)^{1/3}}\left(\frac{M}{\r_0}\right)^{1/3}=0.234\left(\frac{M}{\r_0}\right)^{1/3}. \eeqno{hhuio}
From the results of Ref. \cite{milgrom84} I get for DML isothermal spheres the exact result
\beq r_h=\frac{(2^{1/3}-1)^{2/3}3^{5/3}}{(324\pi)^{1/3}}\left(\frac{M}{\r_0}\right)^{1/3}=0.253\left(\frac{M}{\r_0}\right)^{1/3}.  \eeqno{nityo}
So approximation (\ref{halfash}) remains good for any $n$.
\subsection{The `central-surface-densities relation'}
Reference \cite{milgrom09a} pointed out that MOND predicts a correlation between the central-surface-density of an isolated system, and the same quantity calculated for the `dynamical' mass distribution. The former quantity is the column density of `baryons' -- the true density, $\r(\vr)$ -- along some symmetry axis going through (an assumed) symmetry axis of the system -- such as the symmetry axis of a disc, or any diameter for a spherical system. The second quantity is defined as follows: We first determine the potential field, $\f(\vr)$, of the system, given $\r$ (in our case, using MOND). We then determine the mass distribution, $\r\_D(\vr)$, that would give rise to this potential in Newtonian dynamics:
\beq \r\_D\equiv \frac{1}{4\pi G}\Delta\f. \eeqno{tarew}
A dark-matter adherent would interpret $\r\_D$ as the total, `dynamical' density of baryons plus dark matter. But in MOND, $\r\_D-\r$ is a fictitious, `phantom' density.
\par
Since by MOND, baryons determine the full dynamics, $\r\_D(\vr)$ is calculable from $\r(\vr)$ (in standard dynamics, where dark matter and baryons are separate entities, any values of the two are acceptable as long as $\r\_D\ge\r$).
As corollaries of this general rule, we can derive specific `laws', or correlations, between properties of $\r(\vr)$ and $\r\_D(\vr)$.
\par
Here, I discuss the predicted correlation between the quantities
\beq \S_B\equiv \int\r~dr~~{\rm and}~~~\S\_D\equiv\int\r\_D~dr,  \eeqno{oper}
where the integral is along some symmetry axis.
Note that $\S G$ is a measure of a gravitational acceleration. So it is useful to refer it to $\az$ by defining the `MOND surface density'
\beq \S\_M\equiv \frac{\az}{2\pi G}.   \eeqno{huder}
\par
MOND predicts a strong correlation between $\S\_B$ and $\S\_D$ for the full range of accelerations.
\par
For disc galaxies, Ref. \cite{lelli16} found such a relation between the two attributes evaluated along the symmetry axis. Reference \cite{milgrom16} then showed that modified-gravity formulations of MOND predict an exact, universal such relation that applies to all mass distributions in the disc
\beq \S\_D=\S\_M\SS(\S\_B/\S\_M),  \eeqno{meds}
and showed how $\SS$ is determined from the MOND interpolating function, in very good agreement with the observations.
\par
However, as discussed in Ref. \cite{milgrom09a}, spherical systems do not obey such a universal relation. While MOND still predicts a correlation, the exact relation depends on the mass distribution in the sphere. Here, I use the gamut of DMPs to see how tight such a MOND relation is among them.
\par
For all types of systems, MOND predicts that for $\S\_B\gg\S\_M$, $\S\_D\approx\S\_B$, namely, very little central `phantom matter' for high-surface-density galaxies.
\par
The DML is reached when  $\S\_B\ll\S\_M$.\footnote{$\az$ and $G$ appear in $\S\_M$ not through their product $\azg$. But, there is a $G$ appearing in the definition of $\S\_D$ in terms of the accelerations; so in the end only $\azg$ appears in our DML result.} In this limit MOND predicts an approximate correlation of the form
$\S\_D\propto(\S\_B\S\_M)^{1/2}\gg\S\_B$; so, it is useful to define
\beq \eta\equiv \frac{\S\_D}{(\S\_B\S\_M)^{1/2}},   \eeqno{ssig}
and study how $\eta$ varies among systems.
\par
For the above-mentioned relation for disc galaxies, $\eta=2$ is universal. For spheres, Ref. \cite{milgrom09a} estimated $\eta\sim 2.5$, but variable. So, we now check how variable $\eta$ is among DMPs.
\par
For spheres, in general, we have
$$\S\_D\equiv 2\int\_0\^{\infty}\r\_Ddr\equiv -\frac{1}{2\pi G}\int\_0\^{\infty}\div\vec g dr=$$
\beq -\frac{1}{2\pi G}\int\_0\^{\infty} (g'+2g/r) dr=\frac{1}{\pi G}\int\_0\^{\infty}|g|dr/r, \eeqno{kuop}
where the first term is dropped assuming that the acceleration vanishes at the center as well as at infinity.
\par
For a general spherical mass distribution
\beq \eta=\pi^{-1/2}\frac{\int\_0\^\infty M^{1/2}(r)r^{-2}dr}{[\int\_0\^\infty\r(r) dr]^{1/2}}.   \eeqno{hutdar}
Using the dimensionless variables for our polytropes
\beq \eta=2\frac{\int\_0\^\infty m^{1/2}(y)y^{-2}dy}{[\int\_0\^\infty\z(y) dy]^{1/2}}.   \eeqno{hutes}
This tells us that $\eta$ is independent of the polytropic coefficient $\K$, but that it can depend on $n$.
\par
\begin{figure}[ht]
	\centering
\includegraphics[width = 7cm] {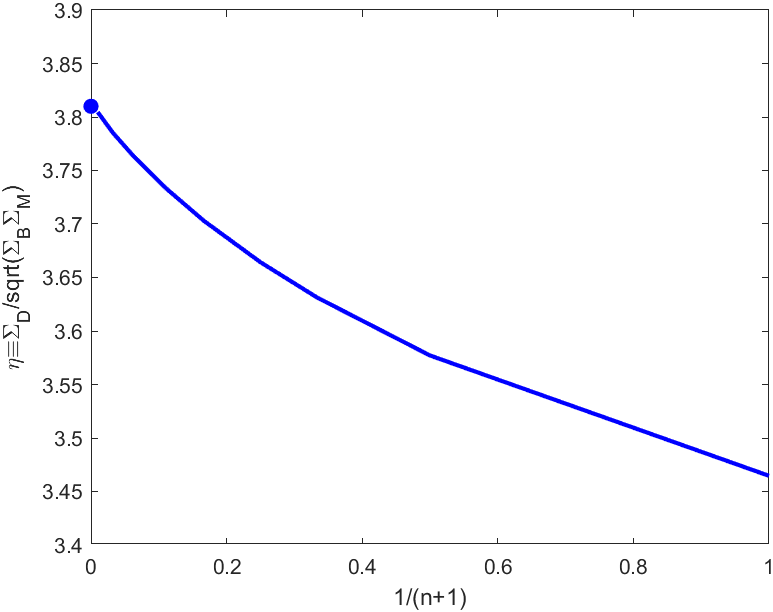}
\caption{The central-surface-densities ratio $\eta$, defined in eq. (\ref{ssig}), vs $(n+1)^{-1}$, for $0\le n\le 100$. Also shown as the dot is the value of $\eta=3.81$ calculated for the isothermal case ($n=\infty$).}
\label{CSDR}
\end{figure}

In Fig. \ref{CSDR}, I show $\eta$ as a function of $n$.
The values of $\eta$ for the extreme cases can be calculated analytically. For $n=0$ (homogeneous sphere), it is $\eta=2\sqrt{3}\approx3.46$.  For $n=\infty$ (isothermal spheres), it can be calculated directly from expression (\ref{uta}) for the density to give $\eta=\sqrt{8\pi/\sqrt{3}}\approx 3.81$,
also shown in Fig. \ref{CSDR}.
We see that while $\eta$ does depend on $n$, it varies only between the above two values for the full range of $n$, exhibiting a rather tight DML central-surface-densities relation.\footnote{For the extreme case of a thin spherical shell we have $\eta=2$.} In Sec. \ref{csdr_anis}, I show the values of $\eta$ for anisotropic polytropes.
\section{Anisotropic-velocity polytropes  \label{anisotropic}}
The class of DMPs with isotropic velocity distribution I discussed above can be enlarged in different ways. There are the spherical polytropes with the full acceleration range from the Newtonian to the DML. One may want to allow for anisotropic velocity dispersions, and one can further generalize by allowing one or more of the EoS parameters ($\K$, $n$, or the anisotropy ratio) to be radius-dependent.
While such generalizations afford larger flexibility, they are also less amenable to the general considerations that underlie our treatment here. For example, allowing $r$ dependence of a parameter introduces a preferred distance scale, which complicates the discussion.
\par
Unlike gas spheres such as stars, where the velocity dispersions are isotropic due to short relaxation times, in stellar systems, such as dwarf spheroidal galaxies, the velocity distributions are known not to be isotropic.
In this Section I thus discuss the more general class of anisotropic polytropes, but with a constant ratio between the tangential and the radial velocity dispersions.
\par
In such systems, the velocity distribution at a point at radius $r$ is characterized by two velocity dispersions, $\s_r$ in the radial direction, and $\s_t$ in any tangential direction. The three-dimensional dispersion is $\s=\sqrt{\s_r^2+2\s_t^2}$, and the anisotropy ratio is defined as $\b=1-\s_t^2/\s_r^2$. The phase-space distribution function of these systems depends on the position and velocity vectors through the energy, $E$, and the angular momentum, $L$, as $f(E,L)\propto L^{-\b}(-E)^{n-3/2}$ .

Defining the pressure $P=\r\s_r^2$, the DML hydrostatic-equilibrium equation (\ref{equilI}) is modified to read
\beq \r\left[\frac{\azg M(r)}{r^2}\right]^{1/2}=-\frac{dP}{dr}-\frac{2\b P}{r}=-r^{-2\b}\frac{d(r^{2\b}P)}{dr}.  \eeqno{equilIV}
These spheres all satisfy the DML, $M-\s$ relation (\ref{poi}); it is a general result of modified-gravity DML, but can also easily be shown following the same derivation as in Sec. \ref{msigma}, {\it mutatis mutandis}.
\par
Using the same units of length and density as in the $\b=0$ case,
 eq. (\ref{equilIV}) is written in the dimensionless form
\beq [m(y)]^{1/2}\equiv \left[\int_0^y \t^n \bar y^2 d\bar y\right]^{1/2}=-y\t'-\xi\t=-y^{1-\xi}(y^\xi\t)',  \eeqno{dimrul}
where
\beq \xi=2\b/(n+1). \eeqno{fyret}
It is useful to work with the dependent variable
$\kappa\equiv y^\xi\t$, in terms of which we have
\beq [m(y)]^{1/2}= \left[\int_0^y \kappa^n \bar y^\eta d\bar y\right]^{1/2}=-y^{1-\xi}\kappa'=-\xi\frac{d\kappa}{d(y^\xi)},  \eeqno{dimbag}
where
\beq \eta\equiv 2-n\xi=2(1-\b/\c). \eeqno{nnuder}
Since $\b\le 1$, we have $\eta\ge 0$ for any $n\ge 0$.
This gives the 2nd order differential equation
\beq \kappa^n(y)=y^{-\eta}[(y^{1-\xi}\kappa')^2]'.  \eeqno{dimopi}
The units of length and density are related, as before, by eq. (\ref{jayu}), in which $n$ and $\K$ appear, but not $\b$.
\par
Note that for $\b\not = 0$, $\t$ is no longer $-\f$, as its derivative is not the MOND acceleration. For this reason, it is possible for $\t$ (and thus $\r$) to increase with radius, because it is not $-\t'$ that balances gravity, but $-y^{-\xi}\kappa'$ that does; so $\kappa$ has to decrease, instead.
\subsection{Behavior at the origin}
We see from eq. (\ref{dimrul}) that for $m(y)$ to vanish at the origin, we cannot have a constant density near the origin. Rather, we have to have $\kappa$ constant there, or
\beq  \t(y)\propto y^{-\xi},  \eeqno{zarut}
or a density $\r\propto r^{-2\b n/(n+1)}=r^{-2\b/\c}$.
As in the isotropic case, a solution with any choice of the normalization -- for example that with $\kappa(0)=1$ [$\t(y)= y^{-\xi}$] and $m(y)=y^{3-\xi}/(3-\xi)$, near the origin -- generates those for all normalizations.
In solving the problem numerically, and presenting the results below, I use eq. (\ref{dimbag}), with the boundary conditions $m(0)=0$ and $\kappa(0)=1$.
\subsection{Behavior near the edge}
The same arguments that led us to deduce that DMPs with $\b=0$ have a finite edge, apply also for $\b\not=0$ (for $\c\not= 1$);
the right-hand side of eq. (\ref{dimbag}) can be written as $-\xi d\kappa/d(y^\xi)$. So as before, $\kappa$ positive that decreases all the way to $\infty$ implies that the right-hand side tends to zero at infinity, inconsistent with the behavior of the left-hand side.
\par
We see from eq.(\ref{dimrul}) that, since $m(y)$ goes to $m_t=m(\yout)$ at $\yout$, the behavior of $\t$ just below $\yout$ has to be
\beq  \t\approx m_t^{1/2}[(y/\yout)^{-\xi}-1]/\xi .  \eeqno{nured}
[This generalizes expression (\ref{logo}) to which it tends for $\b\rar 0$.)
\par
The DML gravitational potential, just below the edge, and everywhere outside it, is still given by eq. (\ref{zalu}), irrespective of the value of $\b$.
\subsection{Analytic solutions}
As in the isotropic case, analytic solutions are known for $n=\infty$ and $n=0$.
The former is the `isothermal' case, for which the analytic solution is given in Ref. \cite{milgrom84}
and reads
\beq  \r(r)=ar^{-2\b}[1+br^{(3-2\b)/2}]^{-3}, \eeqno{hutes}
where $a$ is an arbitrary positive constant, and $b$ is determined from $a$, $\b$, and $\s_r$, so that the total mass, $M$, and the three-dimensional velocity dispersion, $\s$, satisfy the universal DML relation $M\azg=(9/4)\s^4$. It is also the only $n$ value for which the DMP does not have a finite radius.
\par
For $n=0$, eq. (\ref{dimopi}) can be easily solved with the boundary conditions $\kappa(0)$ constant, and $\kappa'(0)=0$:
\beq \kappa=a-\o y^{3/2+\xi};~~~~~\o\equiv 2/\sqrt{3}(3+2\xi), \eeqno{iopik}
where $a$ is any positive constant. Thus
\beq  \t=ay^{-\xi}-\o y^{3/2}.  \eeqno{huter}
The edge, where $t(\yout)=0$ occurs at
\beq \yout=(a/\o)^{2/(3+2\xi)}, \eeqno{jure}
so we can also write
\beq  \t=a\yout^{-\xi}[(y/\yout)^{-\xi}-(y/\yout)^{3/2}].  \eeqno{huter}
\subsection{Numerical results}
\begin{figure}[ht]
	\centering
\includegraphics[width = 7cm] {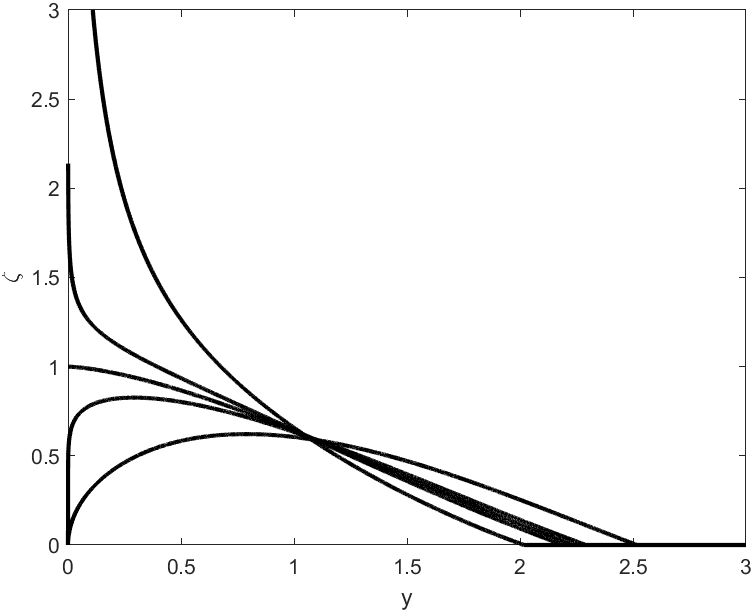}
\caption{The dimensionless density, $\z(y)$, for  $n=1$, and, from bottom to top, $\b=-0.5,~-0.1,~0,~0.1,~0.5$.}		\label{kappa1}
\end{figure}

\begin{figure}[ht]
	\centering
\includegraphics[width = 7cm] {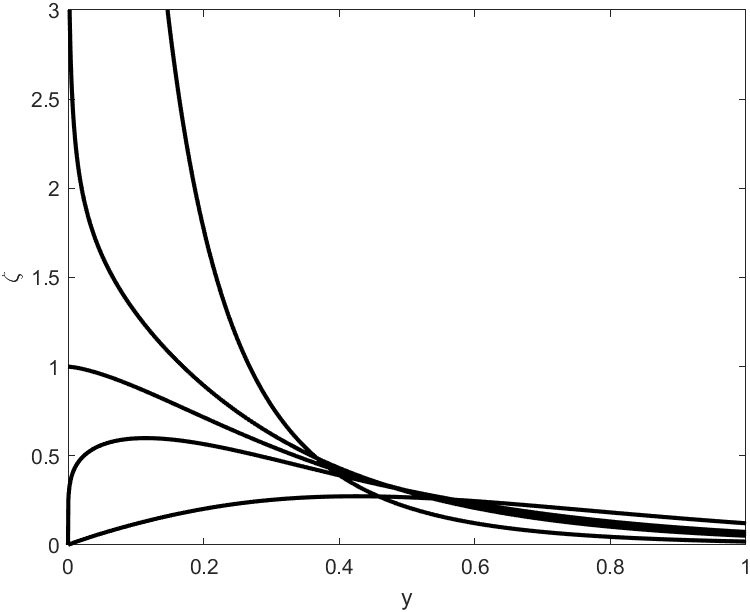}
\caption{The dimensionless density, $\z(y)$, for  $n=10$, and, from bottom to top, $\b=-0.5,~-0.1,~0,~0.1,~0.5$.}		\label{kappa2}
\end{figure}

\begin{figure}[ht]
	\centering
\includegraphics[width = 7cm] {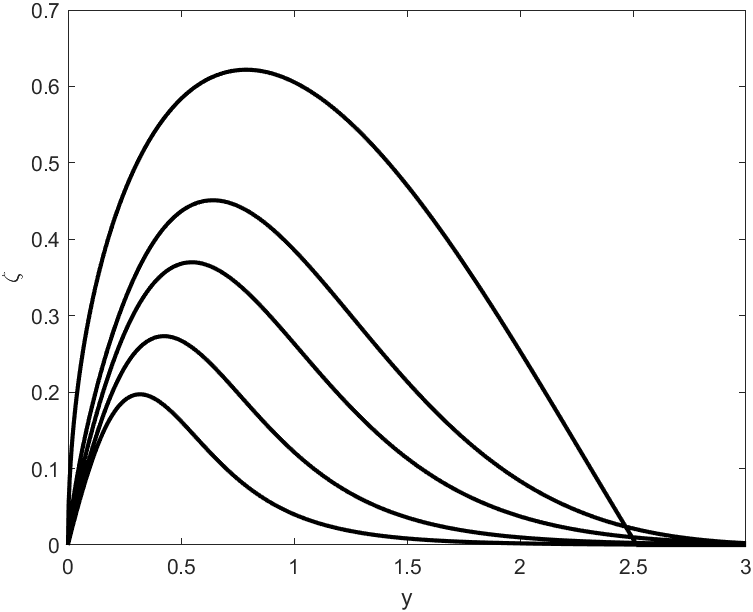}
\caption{The dimensionless density, $\z(y)$, for  $\b=-0.5$ and, from top to bottom, $n=1,~3,~5~10,~20$.}		\label{asa1}
\end{figure}

\begin{figure}[ht]
	\centering
\includegraphics[width = 7cm] {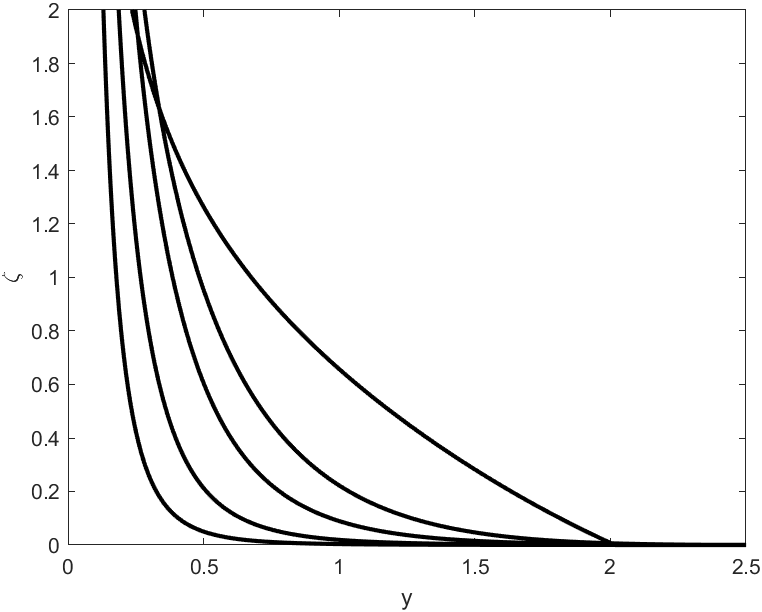}
\caption{The dimensionless density, $\z(y)$, for  $\b=0.5$ and, from top to bottom, $n=1,~3,~5~10,~20$.}		\label{asa2}
\end{figure}

To demonstrate some of the variety afforded by allowing anisotropy, I show some results of numerical solutions using eq. (\ref{dimbag}), with the boundary conditions $m(0)=0$ and $\kappa(0)=1$.
Figure \ref{kappa1} shows the run of the dimensionless density, $\z(y)$, for $n=1$, and $\b=-0.5,~-0.1,~0,~0.1,~0.5$; Fig. \ref{kappa2} shows $\z(y)$ for $n=10$, and the same $\b$ values. Figure \ref{asa1} shows $\z(y)$ for $\b=-0.5$, and $n=1,~3,~5~10,~20$, and Fig. \ref{asa2} shows $\z(y)$ for $\b=0.5$, and the same $n$ values. The last two figures can be compared with Fig. \ref{rhoseries} (or \ref{rhofirst}), which shows $\z$ for $\b=0$ and the same $n$ values.


\subsection{The central-surface-densities relation  \label{csdr_anis}}
\begin{figure}[ht]
	\centering
\includegraphics[width = 7cm] {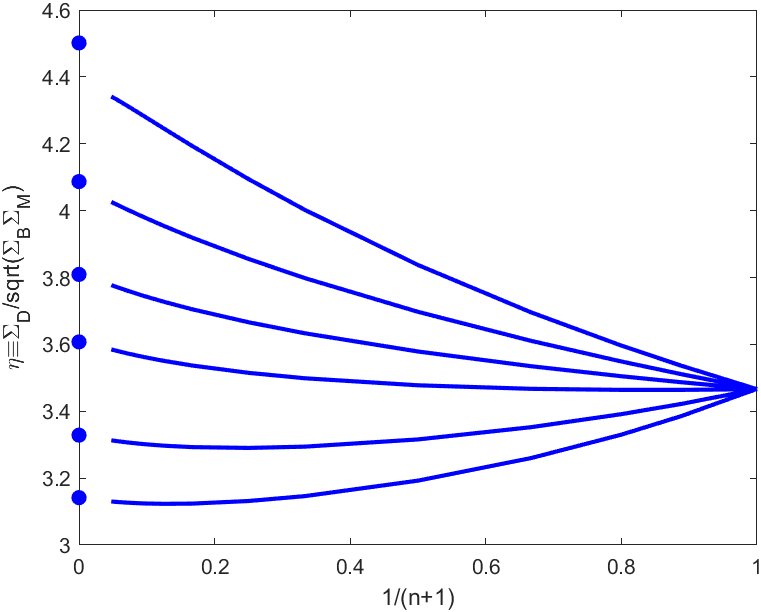}
\caption{The central-surface-densities ratio $\eta$, defined in eq. (\ref{ssig}), vs $(n+1)^{-1}$, for anisotropy ratios (from bottom to top) $\b=-0.5,~-0.3,~-0.1,~0,~0.1,~0.2$. Also shown as the dots are the values of $\eta\_\infty$ calculated analytically for the isothermal case ($n=\infty$), for these $\b$ values, and given in eq. (\ref{katut}).}
\label{CSDRanis}
\end{figure}

In Fig. \ref{CSDRanis}, I show the ratio of central surface densities, $\eta$, defined in eq. (\ref{ssig}), as a function of $n$ for $\b=-0.5,~-0.3,~-0.1,~0,~0.1,~0.2$. Also shown are the limiting values of $\eta$ for `isothermal spheres' ($n=\infty$), calculated analytically from the density law (\ref{hutes}), and found to be
\beq   \eta\_\infty=-\frac{4b^{3/2}\Gamma(-2b)\Gamma(2b)}{\sqrt{\Gamma(4b)\Gamma(2-4b)}}=\frac{\sqrt{4\pi b}}{\sin{(2\pi b)}}\left[\frac{\sin{\pi(1-4b)}}{1-4b}\right]^{1/2},  \eeqno{katut}
where $b=(3-2\b)^{-1}$, and $\Gamma$ is the Gamma function. This result holds only for $\b<1/2$. For $\b\ge1/2$, both $\S\_B$ and $\S\_D$ diverge (see below).
For example, for $\b=-1/2$, $b=1/4$ and $\eta\_\infty=\pi$.
\par
For all $\b$ values, $n=0$ is the homogeneous sphere with $\eta=2\sqrt{3}$.
\par
Another limiting case is $\b\rar -\infty$ for large $n$.
This $\b$ limit corresponds to purely tangential, i.e., circular, orbits of the constituents. Since, for high $n$, the velocities become $r$-independent, the constituents must lie in a single, thin spherical shell. For this case one easily derives $\eta=2$. This checks, as the limit of expression (\ref{katut}) for $b\rar 0$ is 2.
\par
Note that $\b>0$ implies a diverging density at the origin. Such models may still be relevant if we modify the small region around the origin. However, such models may give formally diverging values of $\S\_D$ and $\S\_B$. Near the origin we have $\r\propto r^{-\d}$, where $\d=2\b n/(n+1)$. The MOND acceleration behaves there as $g\propto r^{1-\d}$. Thus, for $\d\ge 1$, both $\S\_B$ and $\S\_D$ diverge. For $\d=1$, they both diverge logarithmically, and so $\eta$ diverges as $\sqrt{\ln{r}}$ at the center; e.g., expression (\ref{katut}) diverges for $\b=1/2$.
For $\d>1$, $\S\_D$ diverges as $r^{(1-\d)/2}$, and $\S\_B$ as $r^{1-\d}$; so the expression for $\eta$ is still formally finite, but it involves a ratio of diverging quantities. Expression (\ref{katut}) for $\eta\_\infty$ is applicable only for $\b<1/2$.
\par
A class of spheres unrelated to polytropes are those with a power-law density distribution, $\r(r)\propto r^k$ within some finite radius and zero outside. Such spheres have $\eta=2\sqrt{(k+3)/(k+1)}$, which span values from $\eta=2\sqrt{3}$ for $k=0$ (homogeneous sphere -- as above), to $\eta=2$ for $k=\infty$, which, again, describes an infinitely thin, hollow, spherical shell. Such thin shells seem to have the lowest $\eta$ value for a spherical distribution.
\par
We see that while $\eta$ does depend on $n$ and $\beta$, the polytropes with the given range of $\b$ and all $n$ values -- and also other spherical systems -- still satisfy a rather narrow DML central-surface-densities relation with $\eta$ varying only by a factor of about 2, between $\eta=2$ and $\eta=4$.

\section{Discussion \label{discussion}}
I have discussed a limited but indicative class of heuristic DML models of self-gravitating spheres, with well-defined distribution functions. These are polytropic spheres, with either isotropic velocity distributions, which I discuss in more detail, or polytropes with a constant velocity anisotropy, treated more succinctly.
\par
One possible enlargement of the class could involve treatment of systems that are not fully in the DML, but the accelerations in which extend from the Newtonian to the DML regime. This was done in Ref. \cite{milgrom84} for only the `isothermal' ($n=\infty$) case. These would be more relevant to `high-surface-density' systems, such as massive elliptical galaxies and globular clusters.
\par
One important thing that we learn about such, more general, MONDian spheres, already from the present study, is that they are all of finite radius (for $\c>1$).
As long as the acceleration stays above $\az$, the sphere behaves as a Lane-Emden one. But then, if the corresponding Lane-Emden sphere is extended, the acceleration must, at large enough radii, enter and remain in the DML. Then our arguments here can be carried to show that the density must drop to 0 at  a finite radius.
\par
For $n<5$, Newtonian polytropes with $MG/r^2\_{out}\gg\az$ are Newtonian everywhere.\footnote{Even for spheres with average accelerations much above $\az$, there may be a small region near the center, where the acceleration drops below $\az$, if $M(r)$ decreases slower than $r^2$ as $r\rar 0$ (e.g., if the density is constant at the center).}


\begin{thebibliography}{}

\bibitem[Milgrom (1983)]{milgrom83}Milgrom M., 1983, A modification of the Newtonian dynamics as a possible alternative to the hidden mass hypothesis. Astrophys. J. 270, 365
\bibitem[Famaey and McGaugh (2012)]{fm12}Famaey, B. and McGaugh, S.S., 2012,  Modified Newtonian Dynamics (MOND): Observational Phenomenology and Relativistic Extensions. Living Rev.  Relativ., 15, 10
\bibitem[Milgrom (2014)]{milgrom14}Milgrom, M., 2014, continually updated,
The MOND paradigm of modified dynamics. Scholarpedia, 9(6), 31410
\bibitem[Milgrom (2019)]{milgrom20}Milgrom, M., 2019, MOND vs. dark matter in light of historical parallels. Stud. Hist. Philos. Mod. Phys. 71, 170
\bibitem[Milgrom (2009)]{milgrom09}Milgrom, M., 2009, The MOND Limit from Spacetime Scale Invariance.
Astrophys. J. 698, 1630
\bibitem[Binney and Tremaine (2008)]{bt08}Binney, J. and Tremain, S. 2008 (2nd Ed.), Galactic Dynamics. Princeton U. Press

\bibitem[Kippenhahn Weigert and Weiss (2012)]{kippen12}Kippenhahn, R., Weigert, A., and Weiss, A., 2012. {\it Stellar Structure and Evolution (2nd Ed.)}. Springer
\bibitem[Milgrom (1984)]{milgrom84}Milgrom M., 1984, Isothermal spheres in the modified dynamics. Astrophys. J. 287, 571

\bibitem[Sanders (2000)]{sanders00}Sanders, R.H., 2000, The fundamental plane of elliptical galaxies with modified Newtonian dynamics. Mon. Not. R. Astron. Soc. 313, 767
\bibitem[Ibata et al. (2011)]{ibata11} Ibata, R., Sollima, A., Nipoti, C., Bellazzini, M., Chapman, S. C., and Dalessandro, E., 2011. Polytropic Model Fits to the Globular Cluster NGC 2419 in Modified Newtonian Dynamics. Astrophys. J. 743, 43
\bibitem[Sanders (2012)]{sanders12}Sanders, R.H., 2012, NGC 2419 does not challenge MOND, Part 2.
    Mon. Not. R. Astron. Soc. Lett. 422, L21
\bibitem[Milgrom (2009a)]{milgrom09a}Milgrom, M., 2009a, The central surface density of `dark haloes' predicted by MOND.
 Mon. Not. R. Astron. Soc. 398, 1023
\bibitem[Lelli et al. (2016)]{lelli16}Lelli, F., McGaugh, S.S., Schombert, J.M., and Pawlowski, M.S. 2016, The Relation between Stellar and Dynamical Surface Densities in the Central Regions of Disk Galaxies. Astrophys. J. Lett. 827, L19
\bibitem[Milgrom (2016)]{milgrom16}Milgrom, M., 2016, Universal MOND relation between the baryonic and `dynamical' central surface densities of disc galaxies. Phys. Rev. Lett. 117, 141101
\bibitem[Bekenstein and Milgrom (1984)]{bm84}Bekenstein, J. and Milgrom, M., 1984, Does the missing mass problem signal the breakdown of Newtonian gravity? Astrophys. J. 286, 7
\bibitem[Milgrom (2010)]{milgrom10}Milgrom, M., 2010, Quasi-linear formulation of MOND. Mon. Not. R. Astron. Soc. 403, 886
\bibitem[Gerhard and Spergel (1992)]{gs92}Gerhard, O.E. and Spergel, D.N., 1992. Dwarf Spheroidal Galaxies and Non-Newtonian Gravity. Astrophys. J. 397, 38
\bibitem[Milgrom (1994)]{milgrom94}Milgrom M., 1994, Modified Dynamics Predictions Agree with Observations of the HI Kinematics in Faint Dwarf Galaxies Contrary to the Conclusions of Lo, Sargent, and Young. Astrophys. J. 429, 540
\bibitem[Milgrom (2014a)]{milgrom14a}Milgrom, M., 2014a, General virial theorem for modified-gravity MOND.  Phy. Rev. D  89, 024016

\end{thebibliography}
\end{document}